\documentclass[review, 5p, times, twocolumn]{elsarticle}

\usepackage{lineno,hyperref}
\usepackage{subfigure}
\usepackage{color}
\usepackage{float}

\journal{Phys. Lett. B}

\begin{document}

\begin{frontmatter}

\title{On the Chiral imbalance and Weibel Instabilities}




\author [prl]{Avdhesh Kumar}
\ead {avdhesh@prl.res.in}

\author [prl]{Jitesh R. Bhatt}
\ead {jeet@prl.res.in}

\author [ipr]{P. K. Kaw}
\ead {kaw@ipr.res.in}

\address [prl]{ Physical Research Laboratory, Navrangpura, Ahmedabad 380 009, India.}
\date{\today}
\address [ipr]{Institute for Plasma Research,
Bhat, Gandhinagar 382428, India.}
\begin{abstract}
We study the chiral-imbalance and the Weibel instabilities in presence of the quantum anomaly using the 
Berry-curvature modified kinetic equation. We argue that in many realistic situations, e.g. 
relativistic heavy-ion collisions, both the instabilities can occur simultaneously.
The Weibel instability depends on the momentum anisotropy parameter
 $\xi$ and the angle ($\theta_n$) between the propagation vector and the anisotropy direction. It has 
maximum growth rate at $\theta_n=0$ while $\theta_n=\pi/2$ corresponds to a damping. 
On the other hand the pure chiral-imbalance instability 
occurs in  an isotropic plasma and depends on difference between the chiral chemical potentials
of right and left-handed particles.
It is shown that when $\theta_n=0$, only for a very 
small values of the anisotropic parameter 
$\xi\sim \xi_c$,  growth rates of the both instabilities are comparable. 
For  the cases  $\xi_c<\xi\ll1$,  $\xi\approx 1$ or $\xi \geq 1$ at $\theta_n=0$,
the Weibel modes dominate over the chiral-imbalance instability if $\mu_5/T\leq1$. However, when 
$\mu_5/T\geq1$, it is possible to have dominance of the chiral-imbalance modes
at certain values of $\theta_n$ for an arbitrary $\xi$. 
\end{abstract}


\end{frontmatter}





The scope of applying kinetic theory to understand variety of many-body problems arising in various branches of
physics is truly enormous \cite{Landau_kinetics}. The conventional Boltzmann or Vlasov equations
imply that the vector current associated with the gauge charges is conserved.
But till recently a very important class of physical phenomena associated with the 
CP-violation or the triangle-anomaly were left out of the purview of a kinetic theory.
In such a phenomenon the axial current is not conserved. 
It should be noted here that there exists a several models of hydrodynamics
which incorporates the effect of CP-violation \cite{Son:2009tf, Banerjee:2012iz, Jensen:2012jy, 
Kharzeev:2010gr}. But a hydrodynamical approach requires that
the system under consideration remains in a thermal and chemical equilibrium.  
However, many  applications of the chiral (CP-violating) physics
may involve a non-equilibrium situation e.g. during the early stages of
relativistic heavy-ion collisions. 
Therefore it is highly desirable to have a proper kinetic theory framework to tackle 
the CP-violating effect. 
Recently there has been a lot of progress in developing such a kinetic theory.
In Ref. \cite{Son:2012wh, Zahed:2012, Stephanov:2012, Chen:2013, Loganayagam:2012, Xiao:2005} 
it was shown that if the Berry curvature\cite{Berry} has nonzero flux across
the Fermi-surface then the particles on the surface can exhibit a chiral anomaly in presence of an external
electromagnetic field. In this formalism  chiral-current $j^\mu$ is not conserved and it can be attributed
to Adler-Bell-Jackiw anomaly \cite{adler69, bell69, Nielsen:1983rb}. It can be shown that
if a system of charged fermions does not conserve parity, it can develop an equilibrium electric current along an
applied external magnetic field \cite{vilenkin:80}. This is so called chiral-magnetic effect (CME). It has been
suggested that a strong magnetic field created in relativistic-heavy-ion experiments
can lead to CME in the quark-gluon plasma \cite{Fukushima:2008xe,Kharzeev08, Warringa08}. 
Indeed the recent experiments with STAR detector at Relativistic Heavy Ion Collider (RHIC)
qualitatively agree with a local parity violation. However,
more investigations are required to attribute this charge asymmetry with the CME \cite{Abelev:2009, Abelev:2010}.
The idea that a Berry-phase can influence the electronic properties [e.g. \cite{Xiao:2010} 
and references cited therein] is well-known in condensed matter literature and it can have applications in 
Weyl semimetal \cite{kim13}, graphene \cite{Sasaki01} etc. 
There exists a deep connection between a CP-violating quantum field theory and 
the kinetic theory with the Berry curvature corrections. In Ref. \cite{Son:2012zy} it was shown
that the parity-odd and parity-even correlations calculated using the modified kinetic theory 
are identical with the perturbative results obtained in next-to-leading order hard dense loop approximation.

In this work we aim to apply the kinetic theory with the Berry curvature 
corrections to some non-equilibrium situations.
 We first note that the results obtained in Refs. \cite{Son:2012wh, Son:2012zy}
are limited to low temperature regime $T\ll\mu_5$, where $\mu_5$ is chiral chemical potential, when
the Fermi surface is well-defined. 
Recently Ref.\cite{Manuel:2013} argues that the domain of validity of the modified kinetic theory 
can be extended beyond the Fermi surface to include the effect of  finite temperature.   
As expected from the considerations
of quantum-field theoretic approach \cite{Itoyama:1983, Liu:1988, Nicola:1994} 
the parity-odd contribution remains temperature independent. 
Recently using the modified-kinetic theory \cite{Son:2012zy} in presence of the chiral imbalance
the collective modes in electromagnetic or quark-gluon plasmas were analyzed \cite{Akamatsu:2013}.
In such a system CP-violating effect can split transverse waves into two branches \cite{Nieves:1989}.
It was found in Ref. \cite{Akamatsu:2013} that in the quasi-static limit i.e. for $\omega \ll k$,
where $\omega$ and $k$ respectively denote frequency and wave-number of the transverse wave,
there exists an unstable mode. The instability can lead to the growth of Chern-Simons number
(or magnetic-helicity in plasma physics parlance) at expense of the chiral imbalance. 
Similar kind of instabilities were found in Refs. 
\cite{Redlich:1985, Rubakov:1985,Joyce:1997,Liane:2005, Boyarsky:2012} 
in different context. 

 It may be possible to observe the instability reported in Ref. \cite{Akamatsu:2013}
in the relativistic heavy-ion collisions. 
But in a realistic scenario 
the initial distribution function $n^0_{\bf p}$ for the strongly interacting
matter formed during the collision can be anisotropic in the momentum space.
This kind of initial distribution known to lead to the Weibel instability
of the transverse modes. In the context of relativistic heavy-ion collision
experiments Weibel instability has been extensively studied 
\cite{Mro:1988,Bhatt:1994,Arnold:2003, Romatschke:2002, romatschke}.
The Weibel instability is also well-known in the condensed matter 
 \cite{Tsitsadze:2009, Hu:1991} and plasma physics literatures 
\cite{Weibel:1959, Fried:1959, Krall} and it can generate magnetic fields in
the plasma. Further it should be emphasized that both the chiral-imbalance and
the Weibel instability can operate in the quasi-static regime. 
 Therefore in the present work we aim to analyze the collective modes in an
anisotropic chiral plasma and study how the chiral-imbalance and Weibel instabilities
can influence each other. 
We believe that the results presented here will be useful in studying Weyl metals and 
the quark-gluon plasma created in relativistic heavy-ion collisions.
We consider weak gauge Field limit and 
assume the following power counting scheme: $\partial_x=O(\delta)$ and $A^{\mu}=O(\epsilon)$. Here,  
$\epsilon$ and $\delta$ are small independent parameters. 
In this senario we use modified collisonless kinetic (Vlasov) equation at the 
leading order in $A^{\mu}$ as given in Ref. \cite{Son:2012zy}:

\begin{equation}
(\partial_{t}+{\mathbf{v}}\cdot{\mathbf{\partial_{x}}})n_{\mathbf{p}}+(e{\mathbf{E}}+e{\mathbf{v}}\times{\mathbf{B}}-
{{\mathbf{\partial}}_{\mathbf{x}}}\mathbf{\epsilon_{p}})\cdot{\mathbf{\partial}}_{\mathbf{p}}n_{\mathbf{p}}=0
\label{kineqwbcc}
\end{equation}
where $\mathbf{v}=\frac{\bf{p}}{p}$, 
$\mathbf{\epsilon_{p}}=p(1-e{\mathbf{B\cdot\Omega_{p}}})$ and 
${\mathbf{\Omega_{p}}}=\pm{\mathbf{p}}/{2 p^{3}}$. Here $\pm$ sign corresponds to right and lefted handed 
fermions respectively. In absence of the Berry curvature term (i.e. $\Omega_{\bf p}$=0) $\epsilon_{\bf{p}}$ is independent of x,   
Eq.(\ref{kineqwbcc}) reduces to the standard Vlasov equation.

\noindent

In this case current density $\bf j$ is defined as:

\begin{eqnarray}
\label{eq:j}
{\bf j} = -e\int \frac{d^3p}{(2\pi)^3}
\Big[\epsilon_{\bf p}{\mathbf{\partial}}_{\mathbf{p}}n_{\mathbf{p}}
+ e\left({\bf \Omega}_{\bf p} \cdot {\mathbf{\partial}}_{\mathbf{p}}n_{\mathbf{p}}\right) \epsilon_{\bf p} {\bf B}
\\ \nonumber + \epsilon_{\bf p} {\bf \Omega}_{\bf p} \times {\mathbf{\partial}}_{\mathbf{x}}n_{\mathbf{p}} \Big]
+e {\bf E} \times {\bf \sigma},
\end{eqnarray}

\noindent where ${\mathbf{\partial}}_{\mathbf{P}}=\frac{\partial}{\partial {\bf p}}$ and 
${\mathbf{\partial}}_{\mathbf{x}}=\frac{\partial}{\partial {\bf x}}$.
The last term on the right hand side of the above equation represents the anomalous Hall current with $\sigma$ given as follows: 
\begin{equation}
{\bf \sigma} = e\int \frac{d^3p}{(2\pi)^3} {\bf \Omega}_{\bf p} n_{\bf p}.
\label{eq:sigma}
\end{equation}

Using Maxwell's equations and linear response theory it is easy to write down the expression for the inverse 
of the propagator in temporal gauge 
$A_0=0$ as follows, 
\begin{equation}
 [({k^2}-{\omega}^{2})\delta^{ij}-k^{i}k^{j}+\Pi^{ij}(K)]E^{j}=[\Delta^{-1}(K)]^{ij}E^{j}=i\omega{j^{i}_{ext}}(k). 
\label{eq1}
\end{equation}
\noindent
Here, $\Pi^{ij}(K)$ is the retarded self energy which follows from expression of the 
induced current $j^{\mu}_{ind}=\Pi^{\mu\nu}(K)A_{\nu}(K)$ and $[\Delta^{-1}(K)]^{ij}$ 
is the inverse of the propagator. Dispersion relation can be obtained by 
finding the poles of the propagator $[\Delta(K)]^{ij}$.

Let us first concentrate on right handed fermions with chemical potential $\mu_{R}$. We consider the background   
distribution of the form $n^{0}_{\mathbf{p}}=1/[e^{(\mathbf{\epsilon_{p}}-\mu_{R})/T}+1]$. In a linear response 
theory we are interested in the induced current by a linear-order deviation 
in the gauge field. We follow  the power counting scheme for gauge field $A_\mu$ and derivatives $\partial_x$ 
as discussed earlier,
and consider deviations in the current and the distribution function up to $O(\epsilon\delta)$. In this 
case we can write the distribution in Eq.(\ref{kineqwbcc}) as follows,
\begin{equation}
n_{\mathbf{p}}=n^{0}_{\mathbf{p}}+e (n^{(\epsilon)}_{\mathbf{p}}+n^{(\epsilon\delta)}_{\mathbf{p}}) 
\label{expansion of distribution function in terms of `epsilon' and `epsilondelta'}
\end{equation}
where, $n^{0}_{\mathbf{p}}$ is the background distribution function in presence of Berry curvature while 
$n^{(\epsilon)}_{\mathbf{p}}$ and $n^{(\epsilon\delta)}_{\mathbf{p}}$ are the pertubations of order 
$O(\epsilon)$ and $O(\epsilon\delta)$ around $n^{0}_{\mathbf{p}}$. Since 
$n^{0}_{\mathbf{p}}$ contains the Berry curvature contribution (Due to $\epsilon_{\mathbf{p}}$) therefore,   
can also be splitted into order $O(0)$ and $O(\epsilon\delta)$ i.e., 
$n^{0}_{\mathbf{p}}=n^{0(0)}_{\mathbf{p}}+e n^{0(\epsilon\delta)}_{\mathbf{p}}$, 
where $n^{0(0)}_{\mathbf{p}}=\frac{1}{[e^{(p-\mu_{R})/T}+1]}$ is the part of background distribution function without 
Berry curvature correction while $n^{0(\epsilon\delta)}_{\mathbf{p}}=\left(\frac{{\mathbf{B}\cdot\mathbf{v}}}{2 p T}\right)
\frac{e^{(p-\mu_{R})/T}}{[e^{(p-\mu_{R})/T}+1]^2}$ is the part of background distribution with Berry curvature correction.
In order to bring in effect of anisotropy we follow the arguments of Ref. \cite{romatschke}. It is assumed that
the anisotropic equilibrium distribution function can be obtained from a spherically symmetric distribution function
by rescaling of one direction in the momentum space. We consider that there is a momentum anisotropy 
in direction of a unit vector ${\mathbf{\hat{n}}}$. Noting that $p=|{\mathbf p}|$, we replace 
$p\rightarrow \sqrt{{\mathbf p}^{2}+\xi({\mathbf p}\cdot{\mathbf{\hat{n}}})^2}$ in the expression of $n^{0}_{\mathbf{p}}$ to get 
anisotropic distribution function. 
Here $\xi$ is an adjustable anisotropy parameter satisfying a condition $\xi>-1$.
It is convenient to define
a new variable  $\tilde{p}$ such that $\tilde{p}=p\sqrt{1+\xi({\mathbf v}\cdot{\mathbf{\hat{n}}})^2}$. Using this 
new variable one can write $n^{0(0)}_{\mathbf{p}}=\frac{1}{[e^{(\tilde{p}-\mu_{R})/T}+1]}$ and  
$n^{0(\epsilon\delta)}_{\mathbf{p}}=\left(\frac{{\mathbf{B}\cdot\mathbf{v}}}{2 \tilde{p} T}\right)
\frac{e^{(\tilde{p}-\mu_{R})/T}}{[e^{(\tilde{p}-\mu_{R})/T}+1]^2}$.

The anomalous Hall current term in Eq.(\ref{eq:j}) can vanish 
if the distribution function is spherically symmetric in the momentum space. However, for an anisotropic
distribution function this may not be true in general. Since the Hall-current term depends on  electric field, 
it can be of order $O(\epsilon\delta)$ or higher.  
As we are interested in finding  deviations in current and distribution function up to order $O (\epsilon\delta)$,
only $n^{0(0)}_{\mathbf{p}}$ would contribute to the Hall current term. 
Next, we consider ${\bf \sigma}$ from Eq.(\ref{eq:sigma}) which can be written as
\begin{equation}
{\bf \sigma}=\frac{e}{2}\int d\Omega d\tilde{p}\frac{\mathbf v}{[1+\xi(\mathbf{v}\cdot\mathbf{\hat{n}})]^{1/2}}
\frac{1}{(1+e^{(\tilde{p}-\mu_{R})/T})}.\label{sigma1}
\end{equation}  
\noindent
Since $\mathbf{v}$ is a unit vector one can express $\mathbf{v}=(sin\theta cos\phi, sin\theta sin\phi, cos\theta)$
in spherical coordinates. By choosing $\mathbf{\hat{n}}$ in $z-$direction, without any loss of generality,
one can have $\mathbf{v}\cdot\mathbf{\hat{n}}=cos\theta$. Thus the angular integral in the above equation
becomes $\int d(cos\theta)d\phi\frac{\mathbf{v}}{(1+\xi cos^2\theta)^{1/2}}$. Therefore 
$\sigma_{x}$ and $\sigma_{y}$ components of Eq.(\ref{sigma1}) will vanish as $\int^{2\pi}_0 sin\phi d\phi=0$ and
$\int^{2\pi}_0 cos\phi d\phi=0$. While $\sigma_{z}$ will vanish because integration with respect to $\cos\theta$ variable 
will yield it ($\sigma_{z}$) to be zero. Thus the anomalous Hall current term will not contribute for the problem at the hand.

Now the kinetic equation (\ref{kineqwbcc}) can be split into two equations valid    
at $O(\epsilon)$ and  $O(\epsilon\delta)$ scales of distribution function as written below, 
\begin{equation}
(\partial_{t}+{\mathbf{v}}\cdot{\mathbf{\partial_{x}}})n^{(\epsilon)}_{\mathbf{p}}=-({\mathbf{E}}+
{\mathbf{v}}\times{\mathbf{B}})\cdot{\mathbf{\partial}}_{\mathbf{p}}n^{0(0)}_{\mathbf{p}}
\label{kEqnfrdistbtnfnaodrepsilon}
\end{equation}
\begin{equation}
(\partial_{t}+{\mathbf{v}}\cdot{\mathbf{\partial_{x}}})(n^{0(\epsilon\delta)}_{\mathbf{p}}+n^{(\epsilon\delta)}_{\mathbf{p}})=
-\frac{1}{e}{{{\mathbf{\partial}}_{\mathbf{x}}}\mathbf{\epsilon_{p}}}\cdot{\mathbf{\partial}}_{\mathbf{p}}{n^{0(0)}_{\mathbf{p}}}
\label{kEqnfrdistbtnfnaodrepsilondelta}
\end{equation}
Equation for the current defined in Eq.(\ref{eq:j}) 
can also split into  $O(\epsilon)$ and $O(\epsilon\delta)$ scales  as given below,
\begin{equation}
 \mathbf{j^{\mu(\epsilon)}}=e^2\int \frac{d^{3}p}{(2\pi)^3}v^{\mu}n^{(\epsilon)}_{\mathbf{p}}
\label{curntwithbccioepsilon}
\end{equation}
\begin{equation}
 \mathbf{j^{i(\epsilon\delta)}}=e^2\int \frac{d^{3}p}{(2\pi)^3}\left[v^{i}n^{(\epsilon\delta)}_{\mathbf{p}}-
 \left(\frac{{v^{j}}}{2 p}\frac{\partial{n^{0(0)}_{\mathbf{p}}}}{\partial{p^{j}}}\right){{B^{i}}}
 -\epsilon^{ijk}\frac{v^{j}}{2 p}
  \frac{\partial{n^{(\epsilon)}_{\mathbf{p}}}}{\partial{x^{k}}}\right]
  \label{curntwithbccioepsilondelta}
\end{equation}
After adding the contribution from all type of species i.e. right/left fermions with charge 
$e$ and chemical potential $\mu_{R}/\mu_{L}$ as well as right/left handed antifermions with 
charge $-e$ and chemical potential $-\mu_{R}/\mu_{L}$, using the expression $j^{\mu}_{ind}=\Pi^{\mu\nu}(K)A_{\nu}(K)$ 
and Eqs. (\ref{kEqnfrdistbtnfnaodrepsilon}, \ref{kEqnfrdistbtnfnaodrepsilondelta}, 
\ref{curntwithbccioepsilon}, \ref{curntwithbccioepsilondelta}) one can obtain the expression for self energy,   
$\Pi^{ij}=\Pi^{ij}_{+}+\Pi^{ij}_{-}$. The expressions for $\Pi^{ij}_{+}$ (parity even part of polarization tensor) and 
$\Pi^{ij}_{-}$ (parity-odd part) can be written as, 
\begin{equation}
 {\Pi^{ij}_{+}(K)}=m^{2}_{D}\int \frac{d\Omega}{4 \pi}
 \frac{{v^{i}}({v^{l}}+\xi(\mathbf{v\cdot{\hat{n}}})\hat{n}^l)}{(1+\xi(\mathbf{v\cdot{\hat{n}}})^2)^2}
 \left(\delta^{jl}+\frac{v^{j}k^{l}}{{\bf{v\cdot k}}+i\epsilon}\right),
\label{selfenergyoepsilon}
\end{equation}
\begin{eqnarray}
 {\Pi^{im}_{-}(K)}=C_{E}\int 
 \frac{d\Omega}{4 \pi}\Bigg[\frac{i{\epsilon}^{jlm}k^{l}v^{j}v^{i}(\omega+
 \xi(\bf{v\cdot\hat{n}})(\bf{k\cdot\hat{n}}))}{({\bf{v\cdot k}}+i{\epsilon})(1+\xi(\mathbf{v\cdot{\hat{n}}})^2)^{3/2}}+\nonumber\\
 \left(\frac{{v^{j}}+\xi(\mathbf{v\cdot{\hat{n}}})\hat{n}^j}
 {(1+\xi(\mathbf{v\cdot{\hat{n}}})^2)^{3/2}}\right){i{\epsilon}^{iml}k^{l}v^{j}}\nonumber\\-{i{\epsilon}^{ijl}k^{l}v^{j}}
 \left(\delta^{mn}+\frac{v^{m}k^{n}}{{\bf{v\cdot k}}+i\epsilon}\right)
 \left(\frac{{v^{n}}+\xi(\mathbf{v\cdot{\hat{n}}})\hat{n}^n}
 {(1+\xi(\mathbf{v\cdot{\hat{n}}})^2)^{3/2}}\right)
 \Bigg] 
\label{selfenergyoepsilondelta}
\end{eqnarray}
where, 
\begin{eqnarray}
 m^{2}_{D}=-\frac{e^2}{2\pi^{2}}\int_{0}^{\infty}d\tilde{p}{\tilde{p}}^{2}
 \Bigg[\frac{\partial{n^{0(0)}_{\mathbf{\tilde{p}}}}(\tilde{p}-\mu_{R})}{\partial{\tilde{p}}}+
 \frac{\partial{n^{0(0)}_{\mathbf{\tilde{p}}}}(\tilde{p}+\mu_{R})}{\partial{\tilde{p}}}\nonumber\\+
 \frac{\partial{n^{0(0)}_{\mathbf{\tilde{p}}}}(\tilde{p}-\mu_{L})}{\partial{\tilde{p}}}+
 \frac{\partial{n^{0(0)}_{\mathbf{\tilde{p}}}}(\tilde{p}-\mu_{L})}{\partial{\tilde{p}}}\Bigg]\nonumber\\
 C_{E}=-\frac{e^2}{4\pi^{2}}\int_{0}^{\infty}d\tilde{p}{\tilde{p}}
 \Bigg[\frac{\partial{n^{0(0)}_{\mathbf{\tilde{p}}}}(\tilde{p}-\mu_{R})}{\partial{\tilde{p}}}-
 \frac{\partial{n^{0(0)}_{\mathbf{\tilde{p}}}}(\tilde{p}+\mu_{R})}{\partial{\tilde{p}}}\nonumber\\-
 \frac{\partial{n^{0(0)}_{\mathbf{\tilde{p}}}}(\tilde{p}-\mu_{L})}{\partial{\tilde{p}}}+
 \frac{\partial{n^{0(0)}_{\mathbf{\tilde{p}}}}(\tilde{p}-\mu_{L})}{\partial{\tilde{p}}}\Bigg]
\label{thermalmass}.
\end{eqnarray}
\noindent
We would like to mention that the total induced current is, $\bf{j}=\bf{j}^{\epsilon}+\bf{j}^{\epsilon\delta}$ where, 
$\bf{j}^{\epsilon}$ gives contribution of the order of the 
square of plasma frequency or $m^2_{D}$. The plasma 
frequency contains additive contribution from the densities of 
all species i.e. right-handed particle/antiparticles and left-handed particles/antiparticles. The 
current $\bf{j}^{\epsilon\delta}$ arises due to chiral imbalance its contribution 
from each plasma specie, depends upon $e\vec{\Omega_{p}}$. 
Since $e\vec{\Omega_{p}}$ can change sign depending on the plasma specie therefore definition of 
$C_{E}$ contains both positive and negative signs. Consequently a relative signs of fermion and anti-fermion are different 
in $m^2_{D}$ and $C_{E}$. After performing above integrations one can 
get $m^{2}_{D}=e^2\left(\frac{\mu_R^2+\mu_{L}^2}{2\pi^2}+\frac{T^2}{3}\right)$ and $C_E=\frac{e^2 \mu_5}{4\pi^2}$, 
where $\mu_5=\mu_R-\mu_{L}$. It should be emphasized here 
that $C_{E}=0$ when there is no chiral imbalance whereas $m^2_{D}\neq0$.
It should be also be noted that the terms with anisotropy parameter $\xi$ are contributing in the parity-odd part 
of the self-energy given by Eq.(\ref{selfenergyoepsilondelta}). Introduction of chemical chemical
potential $\mu_5$ for chiral fermions requires some qualification. Physically the chiral 
chemical potential imply an imbalance between the right handed and left handed fermion. This in turn related to 
the topological charge\cite{Fukushima:2008xe, Redlich:1985}. It should be noted here that due to the axial anomaly 
chiral chemical potential is not associated with any conserved charge. It can still be regarded as `chemical potential' 
if its variation is sufficiently slow\cite{Akamatsu:2013}.

In order to get the expression for the propagator $\Delta^{ij}$ it is necessary to
write $\Pi^{ij}$ in a tensor decomposition. For the present problem we need six independent 
projectors. For an isotropic parity-even plasmas one may need the transverse $P^{ij}_{T}=\delta^{ij}-k^{i}k^{j}/{k^{2}}$
and the longitudinal $P^{ij}_{L}=k^{i}k^{j}/{k^{2}}$ tensor projectors. Due to the presence
{\it{anisotropy}} vector ${\mathbf{\hat{n}}}$ one needs two more projectors $P^{ij}_{n}={\tilde{n}}^i{\tilde{n}}^j/{\tilde{n}}^2$ 
and $P^{ij}_{kn}=k^{i}{\tilde{n}}^{j}+k^{j}{\tilde{n}}^{i}$ \cite{Kobes:1991}.
To account for parity odd effect we have included two anti-symmetric operators $P^{ij}_{A}=i\epsilon^{ijk}{\hat{k}}^k$
and $P^{ij}_{An}=i\epsilon^{ijk}{\tilde{n}}^k$ where, $\tilde{n}^i=(\delta^{ij}-\frac{k^i k^j}{k^2})\hat{n}^j$. 
Thus we write $\Pi^{ij}$ into the basis spanned by the above six operators as:
\begin{equation}
 \Pi^{ij}=\alpha{P}^{ij}_{T}+\beta{P}^{ij}_{L}+\gamma{P}^{ij}_{n}+\delta{P}^{ij}_{kn}+\lambda{P}^{ij}_{A}+\chi{P}^{ij}_{An}
 \label{seexpan}
\end{equation}
where, 
$\alpha$,$\beta$, $\gamma$, $\delta$ $\lambda$ and $\chi$ are 
some scalar functions of $k$ and $\omega$ and are yet to be determined. 
Similarly  we can write $[\Delta^{-1}(k)]^{ij}$ appearing in Eq.(\ref{eq1}) as
\begin{equation}
 [\Delta^{-1}(K)]^{ij}=C_{T}{P}^{ij}_{T}+C_{L}{P}^{ij}_{L}+C_{n}{P}^{ij}_{n}+C_{kn}{P}^{ij}_{kn}+C_{A}{P}^{ij}_{A}+C_{An}{P}^{ij}_{An}.
 \label{invpropexpan}
\end{equation}
Using Eqs.(\ref{eq1}, \ref{seexpan}, \ref{invpropexpan}) one can find relationship between $C$'s and the scalar functions
appearing in Eq.(\ref{seexpan}) as:
\begin{eqnarray}
\nonumber
 C_{T}=k^{2}-{\omega}^{2}+\alpha, 
 C_{L}=-{\omega}^{2}+\beta, 
 C_{n}=\gamma,
 C_{kn}=\delta,\\ 
 C_{A}=\lambda,
 C_{An}=\chi.\label{eq2}
\end{eqnarray} 
For $\xi\rightarrow0$, using Eqs.(\ref{selfenergyoepsilon}-\ref{selfenergyoepsilondelta}) one finds 
$\alpha_{\arrowvert_{\xi=0}}=\Pi_{T}$, 
$\beta_{\arrowvert_{\xi=0}}=\frac{\omega^2}{k^2}\Pi_{L}$, $\gamma_{\arrowvert_{\xi=0}}=0$, $\delta_{\arrowvert_{\xi=0}}=0$, 
$\lambda_{\arrowvert_{\xi=0}}=-\frac{\Pi_{A}}{2}$ and $\chi_{\arrowvert_{\xi=0}}=0$ where,
\begin{eqnarray}
\Pi_{T}=m^{2}_{D}\frac{\omega^{2}}{2 k^{2}}\left[1+\frac{k^{2}-{\omega}^{2}}{2\omega k}\ln\frac{\omega+k}{\omega-k}\right],\nonumber\\
\Pi_{L}=m^{2}_{D}\left[\frac{\omega}{2 k}\ln\frac{\omega+k}{\omega-k}-1\right],\nonumber\\
\Pi_{A}=-2 k C_{E}\left(1-\frac{{\omega}^2}{k^2}\right)\left[1-\frac{\omega}{2 k}\ln\frac{\omega+k}{\omega-k}\right].
\label{isotropic:eq}
\end{eqnarray}
Scalar functions $\Pi_T$, $\Pi_{L}$ and $\Pi_A$ respectively describe the transverse, longitudinal
and the axial parts of the self-energy decomposition when $\xi=0$\cite{Akamatsu:2013}. 

Using the orthogonality condition, 
$[\Delta^{-1}(K)]^{ij} [\Delta(K)]^{jl}=\delta^{il}$, $[\Delta(K)]^{jl}$ can be determined. Poles of $[\Delta(K)]^{jl}$ 
are given by following equation.
\begin{eqnarray}
2 k {\tilde{n}}^{2}C_{A}C_{An}C_{kn}+C^{2}_{A}C_{L}+{\tilde{n}}^{2}C^{2}_{An}(C_{n}
+C_{T})\nonumber\\+C_{T}(k^{2}{\tilde{n}}^{2}C^{2}_{kn}-C_{L}(C_{n}+C_{T}))=0.~~~
\label{dispersionrelation}
\end{eqnarray}
Eq.(\ref{dispersionrelation}) is the general dispersion relation and 
it is quite complicated to solve analytically or numerically.
Here we would like to ascertain that $\alpha$,
$\beta$, $\gamma$ and $\delta$ appearing in C's are same as those given in Ref. \cite{romatschke}. 
The new contributions come in terms of  $\lambda$ and $\chi$ which contain the
effect of parity violation.  
The standard criterion for the  Weibel instability 
 \cite{Arnold:2003} is not applicable here due to the parity violating effect.
 First we note that the chiral instability occurs in the quasi-stationary regime i.e $|\omega| \ll k$
and if the initial distribution function of the plasma is isotropic then the chiral modes
have an isotropic dispersion relation. While the Weibel instability can occur due to an anisotropy in 
the initial momentum distribution in the plasma and the instability can be present in the quasi-stationary regime.
We study numerical solutions of Eq.(\ref{dispersionrelation}) in quasi-stationary limit.
Further we note that when $C_A, C_{An}=0$, there is no chiral-imbalance and one can get the pure Weibel modes
from Eq.(\ref{dispersionrelation}). The pure chiral-imbalance modes can be obtained by setting $C_n, C_{kn}, C_{An}=0$
in Eq.(\ref{dispersionrelation}).
In order to obtain the growth-rates for the instabilities, one needs to solve Eq.(\ref{dispersionrelation}) for $\omega$.
By setting $\frac{\partial \omega}{\partial k}=0$ one can find $k_{max}$ for which the instability
can grow maximally. Upon substituting  $k_{max}$ in the expression for $\omega$ and using $\omega=i\Gamma$, one
can find the growth rate $\Gamma$ for the instability.

Figs.(1-2) depicts 
a comparison between the  pure Weibel modes (i.e.  $\mu_5=0$) with the mixed modes i.e. when 
both chiral-imbalance and momentum-anisotropy are present.
 Before we discus the result, it should be noted that direction between the propagation vector ${\bf k}$
 and the anisotropy vector ${\hat{\bf n}}$ is quantified by angle $\theta_n$ i.e. 
 ${\bf k}\cdot{\hat{\bf n}}=k\cos{\theta_n}$ where, $k$ is magnitude of vector ${\bf k}$. 
 In Figs. (1a-1b) we have considered the case $\theta_n=0$ at $\mu_5/T=1$ and $\mu_5/T=10$ for the mixed 
 modes respectively; while, $\mu_5/T=0$ is for pure Weibel modes.  These figures show that
 the Weibel modes become strong with increasing values of anisotropy parameter $\xi$. It can also be seen 
 that by increasing $\mu_5/T$ the chiral-modes become stronger, leading to enhancement of mixed modes. In 
 the discussion below we have obtained analytic results 
 for $\xi\ll 1$ and found a critical value $\xi_c$ at $\theta_n=0$ such that for $\xi<\xi_c$ the chiral modes will dominate
 while for $\xi>\xi_c$ the Weibel instability can dominate. Fig.(2) depicts the case when $\theta_n=\pi/2$.
 Here as it is well-known the pure Weibel modes are damped. The damping is increasing with increasing
 $\xi$ but it can become weaker by increasing $\mu_5/T$.

\begin{figure}[H]
\begin{center}
\subfigure[]{\includegraphics[bb=0 0 433 181,width=0.45\textwidth]{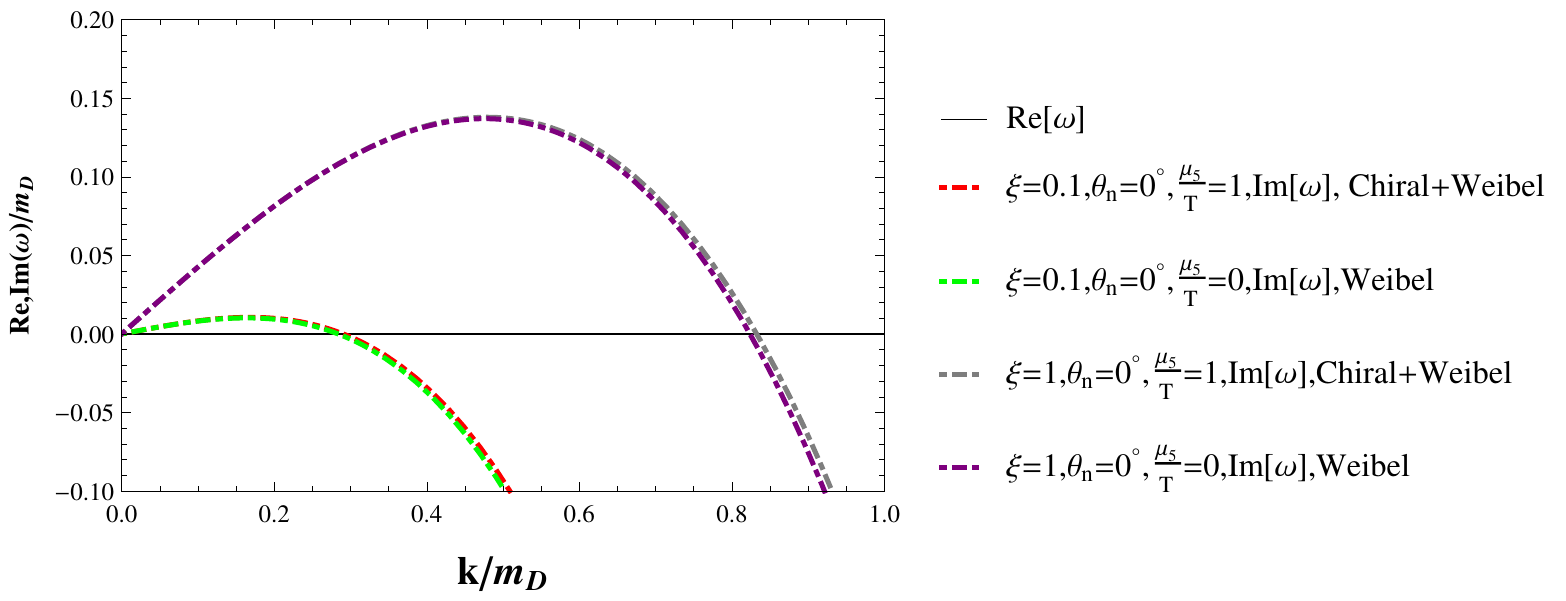}}
\subfigure[]{\includegraphics[bb=0 0 433 181,width=0.45\textwidth]{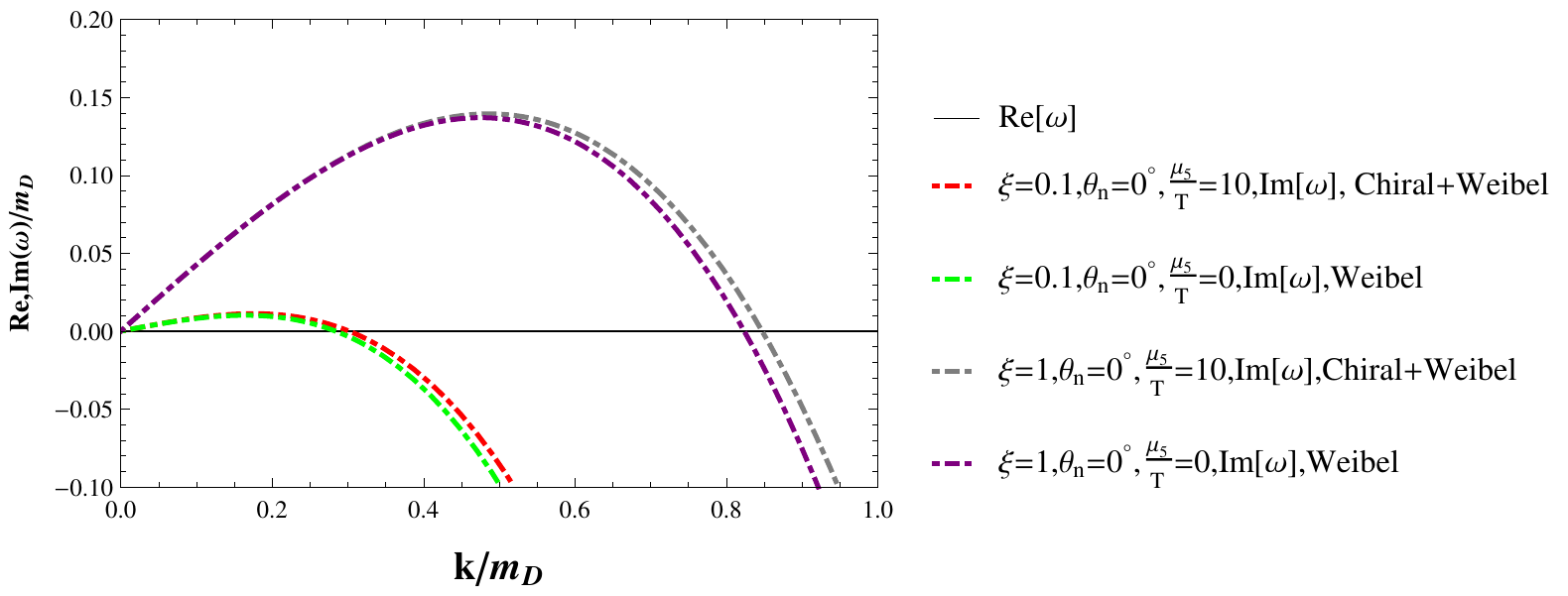}}
 \caption 
 {\label{Fig[1]} 
 Shows plots of real and imaginary part of the transverse dispersion relation for the case when 
 the angle $\theta_n$ between the propagation vector ${\bf k}$ of the perturbation and the anisotropy
 direction is zero.  The modes are purely imaginary and the real part of frequency $\omega=0$.
 Fig. (1a) shows comparison between pure Weibel modes ($\mu_5$=0) with the cases when both the Weibel
 and chiral-imbalance instabilities are present when $\mu_5/T=$1 and $\xi=$0.1,1 .
 Fig. (1b) depicts the similar comparison when $\mu_5/T=$10. It shows that by increasing $\mu_5/T$
 the chiral-imbalance instability become stronger.}
\label{fig[1]}
\end{center}
\end{figure}
\begin{figure}[H]
\begin{center}
{\includegraphics[bb=0 0 433 181,width=0.45\textwidth]{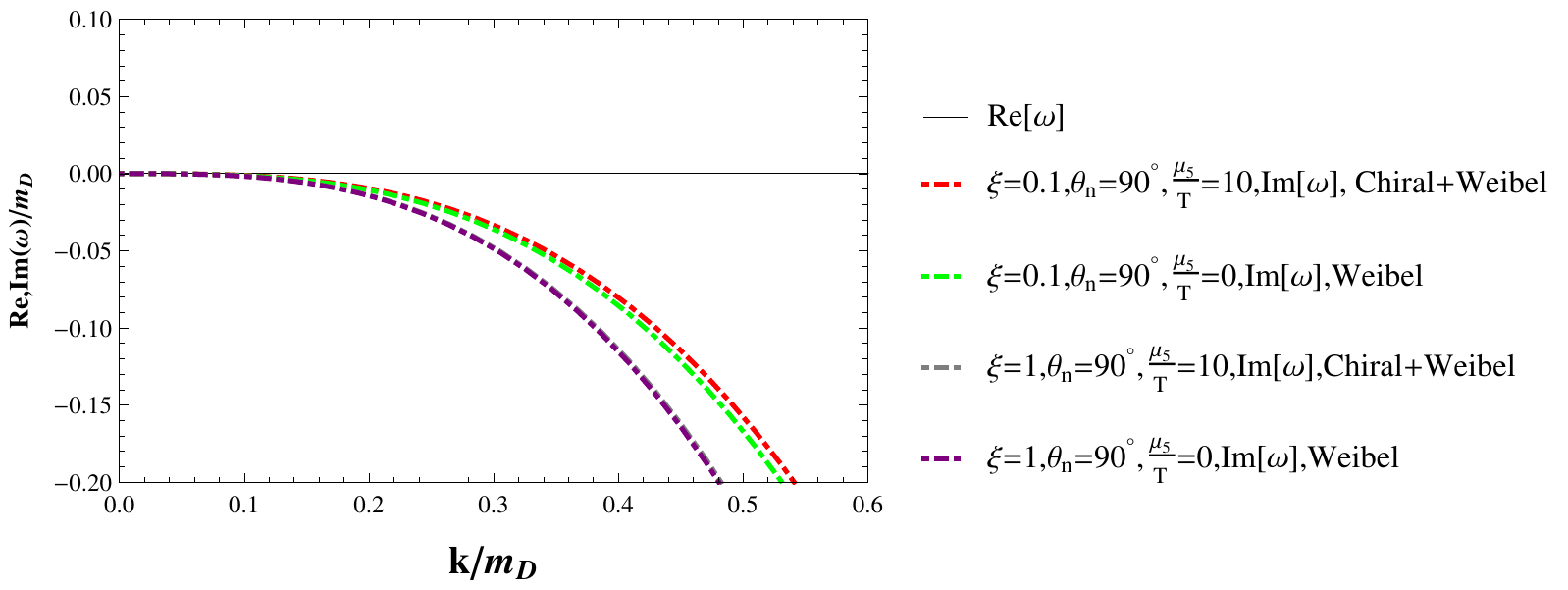}}
 \caption 
 {\label{fig[2]} Shows plots of the dispersion relation when $\theta_n=\pi/2$.
 The pure Weibel modes are known to give damping when $\theta_n=\pi/2$. 
 For the instances 
  when both the chiral-imbalance and Weibel instabilities are present ( $\mu_5/T=$10 and $\xi=$ 0.1,1) 
 the damping can become weaker.} 
\label{fig[2]}
\end{center}
\end{figure}

It is important to notice that there also exists a situation $\xi>>1$ when the chiral-imbalance instability
can play a dominant role in anisotropic plasma. This is because
the Weibel instability growth rate is dependent on $\theta_n$ and 
it is possible to find a particular value of $\theta_n=\theta_{nc}$ 
when the growth rate of the pure-Weibel mode is close to zero. By setting $\omega=0$ in the pure Weibel dispersion
relation, 
one can find for $\xi>>1$, $\theta_{nc}\sim \left(\frac{\pi m_D^2}{2 k^2}\right)^{1/2}\frac{1}{\xi^{1/4}}$.  
In the regime $\xi< 1$ but closer to unity at $\theta_n=0$, 
a comparison between the growth rates of the 
chiral-imbalance ($\Gamma_{ch}$) and Weibel ($\Gamma_{w}$) instabilities is given in the following table:
\begin{center}
\begin{tabular}{ |c|c|c|c|c|c| } 
 \hline
 $\xi$  & 0.6 & 0.7 & 0.8 & 0.9 \\
 \hline
 $\frac{\Gamma_{ch}}{\Gamma_w}$ & $\frac{0.0088 \alpha_e^{3/2} \mu_5^3}{T^3}$ 
 & $\frac{0.0076 \alpha_e^{3/2} \mu_5^3}{T^3}$ & $\frac{0.0067 \alpha_e^{3/2} 
 \mu_5^3}{T^3}$ & $\frac{0.0060 \alpha_e^{3/2} \mu_5^3}{T^3}$\\ 
 \hline
\end{tabular}
\end{center}
Thus the ratio $\frac{\Gamma_{ch}}{\Gamma_w}$ decreases by increasing  values $\xi$ while keeping 
$\mu_5/T$ fixed. This is because $\Gamma_{w}$ is increases by increasing $\xi$. For $\alpha_e=\frac{1}{137}$ and 
$\mu_5/T\leq1$ one can clearly see from the table that the ratio $\frac{\Gamma_{ch}}{\Gamma_w}\ll1$. Thus 
Weibel modes dominates in this case. However when $\mu_5/T\gg1$ chiral 
modes can also dominate.  

Now we consider the case $\xi\ll 1$. This approximation is valid when the initial
momentum anisotropy is weak or the Weibel instability has already nearly thermalized (or isotropized) the plasma.
This may not be an unlikely scenario in the heavy-ion collisions as the growth rates for the Weibel instabilities 
can be much larger than the chiral instability. In this case it is possible to evaluate all the
integrals in the dispersion relation analytically and one can 
express $\alpha$, $\beta$, $\gamma$, $\delta$, $\lambda$ and $\chi$
up to linear order in $\xi$ as follows,
\begin{eqnarray}
\alpha=\Pi_{T}+\xi\Big[\frac{z^2}{12}(3+5\cos{2\theta_{n}})m^{2}_{D}-\frac{1}{6}(1+\cos{2\theta_{n}})m^{2}_{D}\nonumber\\~~~~+
\frac{1}{4}\Pi_{T}\left((1+3\cos{2\theta_{n}})-z^{2}(3+5\cos2\theta_{n})\right)\Big];\nonumber\\
z^{-2}\beta=\Pi_{L}+\xi\Big[\frac{1}{6}(1+3\cos{2\theta_{n}})m^{2}_{D}+\Pi_{L}
\Big(\cos{2\theta_{n}}-\nonumber\\ \frac{z^2}{2}(1+3\cos{2\theta_{n}})\Big)\Big];\nonumber\\
\gamma=\frac{\xi}{3}(3\Pi_{T}-m^{2}_{D})(z^2-1)\sin^2{\theta_{n}};\nonumber\\
\delta=\frac{\xi}{3 k}(4 z^{2} m^{2}_{D}+3\Pi_{T}(1-4z^{2}))\cos{\theta_{n}};\nonumber\\
\lambda=-\frac{\mu_5 k e^2}{4\pi^2}\Big[(1-z^2)\frac{\Pi_{L}}{m^{2}_{D}}\Big]-\xi\frac{\mu_5 k e^2}{32 \pi^2}
\Bigg[(1-z^2)\frac{\Pi_{L}}{m^{2}_{D}}\times\nonumber\\ \left((1+7\cos{2\theta_{n}})-3z^2(1+3\cos{2\theta_{n}})\right)\nonumber\\+
\frac{1}{3}(1+11\cos{2\theta_{n}})-z^2(3+5\cos{2\theta_{n}})
\Bigg];\nonumber\\
\chi=\xi\left[f(\omega,k)\right],~~~~~\label{eq3}
\end{eqnarray}
\noindent
where $z=\frac{\omega}{k}$ and $f(\omega,k)$ is some function $k$ and $\omega$. But in the present analysis 
exact form of $f(\omega,k)$ may not be required.
Using the above equations and Eqs. (\ref{eq2}, \ref{isotropic:eq}) one can finally express
Eq.(\ref{dispersionrelation})  in terms of $k$ and $\omega$. 
One can notice from Eq.(\ref{eq3}) that the most significant contribution for  
$\gamma$, $\delta$, $\lambda$ and $\chi$ is $O(\xi)$. Thus in the present scheme of 
approximation one can write Eq.(\ref{dispersionrelation}) up to $O(\xi)$ as: 
\begin{equation}
 C^{2}_{A}C_{L}-C_{T}C_{L}(C_{n}+C_{T})=0, \label{dispersionrelation1}
\end{equation}
which in turn can give following two branches of the dispersion relation,
\begin{eqnarray}
\label{dispersionrelation2}
C^{2}_{A}-C^{2}_{T}-C_{n}C_{T}=0,\\
C_{L}=0.\label{dispersionrelation3}
\end{eqnarray}
First, we would like to note that when $C_{A}=0$, Eqs.(\ref{dispersionrelation2}-\ref{dispersionrelation3}) reduces to exactly the same
dispersion relation discussed in Ref.\cite{romatschke} for the Weibel instability in an anisotropic plasma when
there is no parity violating effect.
Let us consider Eq.(\ref{dispersionrelation2}), it can be written as:
\begin{equation}
(k^2-\omega^2)^2+(k^2-\omega^2)(2\alpha+\gamma)+\alpha^2+\alpha\gamma-\lambda^2=0.
\end{equation}
This equation is a quadratic equation in $(k^2-\omega^2)$ and it's solutions can be written as,
\begin{equation}
(k^2-\omega^2)=\frac{-(2\alpha+\gamma)\pm2\lambda}{2}.\label{eq4}
\end{equation}

Now, it is of particular interest to consider 
the quasi-static limit $\left|{\omega}\right|<<k$,
in this limit expressions for $\alpha\sim\Pi_T$ and 
$\beta\sim\frac{\omega^2}{k^2}\Pi_L$ and $\lambda\sim-\frac{\Pi_A}{2}$. Now $\Pi_L$, $\Pi_T$ and $\Pi_A$ can be obtained 
by expanding Eq.(\ref{isotropic:eq}) in the quasi static limit as:
\begin{eqnarray}
{\Pi_{T}}_{\arrowvert_{\left|{\omega}\right|<<k}}=\left(\mp{i}\frac{\pi}{4}\frac{\omega}{k}\right)m^{2}_{D};\nonumber\\
{\Pi_{L}}_{\arrowvert_{\left|{\omega}\right|<<k}}= m^{2}_{D}\left[\mp{i}\frac{\pi}{2}\frac{\omega}{k}-1\right]\nonumber\\
{\Pi_{A}}_{\arrowvert_{\left|{\omega}\right|<<k}}=
\frac{\mu_5 k e^2}{2 \pi^2}\left(\frac{{\Pi_{L}}_{\arrowvert_{\left|{\omega}\right|<<k}}}{m^2_D}\right)\label{quasilimitofpit}
\end{eqnarray}
Thus in the quasi-stationary limit one can write positive branch of the transverse modes given
by Eq.(\ref{eq4}) as:
\begin{equation}
\rho(k)=\frac{\left(\frac{4\alpha_{e}\mu_5}{\pi^{2}m^2_{D}}\right)k^2\left[1-\frac{\pi k}{\mu_5\alpha_e}+
\frac{\xi\left(1+5\cos{2\theta_{n}}\right)}{12}
+\frac{\xi\left(1+3\cos{2\theta_{n}}\right)}{12}\frac{\pi m^{2}_{D}}{\mu_5\alpha_{e}k}\right]}
{{\left[1+\frac{2 \mu_5 \alpha_{e} k}{\pi m^2_{D}}(1-\frac{\xi}{4})+\xi\cos{2\theta_{n}}\left(1-
\frac{7 \mu_5 \alpha_{e} k}{2 \pi m^2_{D}}\right)\right]}}
\label{eq5} 
\end{equation}
Here we have used $\omega=i\rho(k)$ and defined $\alpha_{e}=\frac{e^2}{4\pi}$ as the electromagnetic coupling. 
It is clear from Eq.(\ref{eq5}) that $\omega$ is purely an imaginary number and its real-part is zero i.e. $Re(\omega)=0$. 
Positive  $\rho(k)>0$ implies an instability as $e^{-i(i\rho(k))t}\sim e^{+\rho(k)t}$. 
From Eq.(\ref{eq5}), in the limit $\xi\rightarrow0$ and $\mu_5\rightarrow0$
one gets $\rho(k)=-\frac{4k^3}{\pi m^2_D}$. Thus for an isotropic plasma (of massless particles) without any 
chiral-imbalance there is no unstable propagating mode  when $\omega\ll k$. Which is consistent with fact that without 
any source of free energy there should not be any unstable mode. 

Now let us first consider that the quasi-static limit $|\omega|<<k$ indeed satisfies for Eq.(\ref{eq5}). Since 
we have already assumed that $\xi<<1$ and $\alpha_{e}<1$ and for $\mu_5<<T$ one can have 
$\frac{\mu_5}{m_D}\approx\frac{1}{2 \alpha_{e}^{1/2}}\left(\frac{\mu_5}{T}\right)$. From this it is rather easy to 
show that $\rho/k<<1$ if the condition  
$\frac{k^2}{m_{D}^2}<<1$ is satisfied. In this case denominator of Eq.(\ref{eq5}) can be approximated to unity. Now 
we write the above equation as: 
{\small
\begin{eqnarray}
\rho(k)=\frac{4}{\pi}\frac{k^2}{m^2_{D}}\Bigg[\frac{\alpha_{e}\mu_5}{\pi}-k+
\frac{\alpha_e \xi \mu_5}{12 \pi}\left(1+5\cos{2\theta_{n}}\right)
+\nonumber \\ \frac{\xi}{12}\frac{m^{2}_{D}}{k}\left(1+3\cos{2\theta_{n}}\right)\Bigg].
\label{result:1}
\end{eqnarray}}
Here we emphasize that when $\xi=0$, first two terms in the square bracket survives and Eq.(\ref{result:1}) 
matches with the dispersion relation of the chiral instability 
given in Ref.\cite{Akamatsu:2013} and when $\mu_5=0$, the second and the last term survives to  
give the Weibel modes considered in Ref.\cite{romatschke}. Term with $\alpha_{e} \xi \mu_5$ factor arises due 
to the interaction between the Weibel and chiral-imbalance modes.

Before we analyse the interplay between the chiral-imbalance and the Weibel instabilities, it is instructive to qualitatively
understand their origin.  First consider the chiral-imbalance instability. For a such a plasma `chiral-charge' density $n$
is given by $\partial_t n+{\bf  \nabla \cdot j}=\frac{2\alpha_{e}} {\pi} {\bf E \cdot B} $.  From this 
one can estimate the axial charge density 
$n\sim \alpha_{e} k A^2$ where $A$ is the gauge-field. 
Assuming that there are only right handed 
particles i.e ($\mu_5\sim\mu_R$) then the number and energy densities of the plasma respectively 
given by $\mu_5 T^2$ and ${\mu_5}^2 T^2$.
The fermionic number density associated with the gauge field can be estimated from the Chern-Simon term to be
 $\alpha_{e} kA^2$. The
number densities associated with the fields and particles have same value
for $k_1\sim \frac{\mu_5 T^2}{\alpha_{e} A^2}$. The 
typical energy for the gauge field $\epsilon_A\sim k^2A^2$. For this particular value of $k_1$
it can be seen that $\epsilon_A={\mu_5}^2 T^2\frac{T^2}{\alpha_{e}^2 A^2}$. Thus there
exists a state satisfying the condition $\frac{T^2}{\alpha_{e}^2} <A^2$ for which 
energy in the gauge field is lower than  particle energy. 
This leads to the chiral-imbalance instability\cite{Akamatsu:2013,Joyce:1997}.
The Weibel instability arises when the equilibrium distribution function of the plasma has anisotropy in the momentum 
space\cite{Weibel:1959, Fried:1959}.
The anisotropy  in the momentum space can be regarded as anisotropy in temperature. Suppose there is plasma which is 
hotter in $y$-direction than $x$ or $z$ direction one may write the distribution 
function $ n^0_{p}=\frac{1}{1+e^{-(\sqrt{p^2_{x}+(1+\xi)p^2_{y}+p^2_{z}})/T}}$ . If in this situation a 
disturbance with a magnetic-field $B=B_0cos(k x)$ which 
arises  say from noise,  one can write the Lorentz force term in the kinetic equation as 
$e (v\times B)\cdot\partial_{p}n^0_{p}=
e[\xi(v_{z}B_{x}-v_{x}B_{z})\frac{p_{y}}{T}]\left(\frac{-e^{-(\sqrt{p^2_{x}+
(1+\xi)p^2_{y}+p^2_{z}})/T}}{1+e^{-(\sqrt{p^2_{x}+(1+\xi)p^2_{y}+p^2_{z}})/T}}\right)$.
This Lorentz-force can produce current-sheets where the magnetic field changes its sign.
The current-sheet in turn enhance the original magnetic field \cite{Weibel:1959, Fried:1959}. 

The Weibel instability is known to grow maximally for $\theta_n=0$. In 
the quasi-static limit the instability has maximum growth rate $\Gamma_w\sim\frac{8 \xi^{3/2}}{27 \pi}m_D$ for 
$k=\frac{\sqrt{\xi}}{3}m_{D}$. 
For the chiral imbalance instability the maximum growth 
rates $\Gamma_{ch}\sim\frac{16 \alpha_{e}^3}{27 \pi^4}\left(\frac{\mu_5}{m_D}\right)^2\mu_{5}$,  
occurs at $k\sim\frac{2\alpha_{e}}{3\pi}\mu_5$\cite{Akamatsu:2013}. Thus the ratio 
$\frac{\Gamma_{ch}}{\Gamma_{w}}\sim\frac{2}{\pi^3}\left(\frac{\alpha_{e}}{\xi^{1/2}}\right)^3\left(\frac{\mu_5}{m_{D}}\right)^3$ 
$\sim\frac{1}{4\pi^3}\left(\frac{\alpha_{e}}{\xi}\right)^{3/2}\left(\frac{\mu_5}{T}\right)^3$, where we have used 
$\frac{\mu_5}{m_D}\approx\frac{1}{2 \alpha_{e}^{1/2}}\left(\frac{\mu_5}{T}\right)$. The ratio $\frac{\Gamma_{ch}}{\Gamma_{w}}$ 
becomes unity when $\xi_{c}\approx2^{2/3}\left(\frac{\alpha_{e}}{4\pi^2}\right)\left(\frac{\mu_5}{T}\right)^2$. When 
$\mu_5\sim T$  and $\alpha_{e}=1/137$ (QED) one can estimate 
$\xi_c<10^{-3}$. $\xi_c$ will change if coupling varries (QCD case). Thus for $\xi_c<\xi<<1$ the 
Weibel instability 
can dominates over the chiral imbalance modes. However, it may be still possible to see the chiral-imbalance modes if we 
consider $\theta_{n}-$dependence of the instability described by Eq.(\ref{result:1}). In Eq.(\ref{result:1}) the Weibel instability 
term vanishes if $\theta_{n}\sim\frac{1}{2}\cos^{-1}(1/3)\sim55^{\circ}$. For this value of $\theta_{n}$ the interaction 
term between the Weibel and the 
chiral modes becomes negative and tries to suppress the unstable mode. However this term is very small 
in comparison to the pure chiral term. 

In figure (3) we plot the dispersion relation given by Eq.(\ref{eq5}) as function
of $k_N=\frac{\pi}{\alpha_e\mu_5}k$ for various values of $\xi$ which is given in units of $\xi_c$ and 
the propagation angle $\theta_{n}$. $y$-axis
shows the $Re[\omega]$ and $Im[\omega]/\left(\frac{4\alpha_{e}^3{\mu_5}^3}{\pi^{4}m^2_{D}}\right)$. Note that the real 
part of the frequency  $Re[\omega]$ is zero 
For the case when $\xi=0$ there is no Weibel mode
and the only the chiral-imbalance can give the instability. Whereas when $\mu_5=0$ only Weibel
instability will contribute. From the condition $\rho(k)>0$, one can obtain the range of the
instability which can be stated as: 
{\small
\begin{equation}
k_{N}=1+\frac{\xi\left(1+cos2\theta_n\right)}{12}
+\Bigg[\left( 1+\frac{\xi\left(1+cos2\theta_n\right)}{12}\right)^2+
\frac{\pi^2\xi\left(1+3cos2\theta_n\right)}{3\alpha_{e}}\Bigg]^{1/2}
\end{equation}}
\begin{figure}[H]
\begin{center}
\subfigure[]{\includegraphics[bb=0 0 411 181,width=0.45\textwidth]{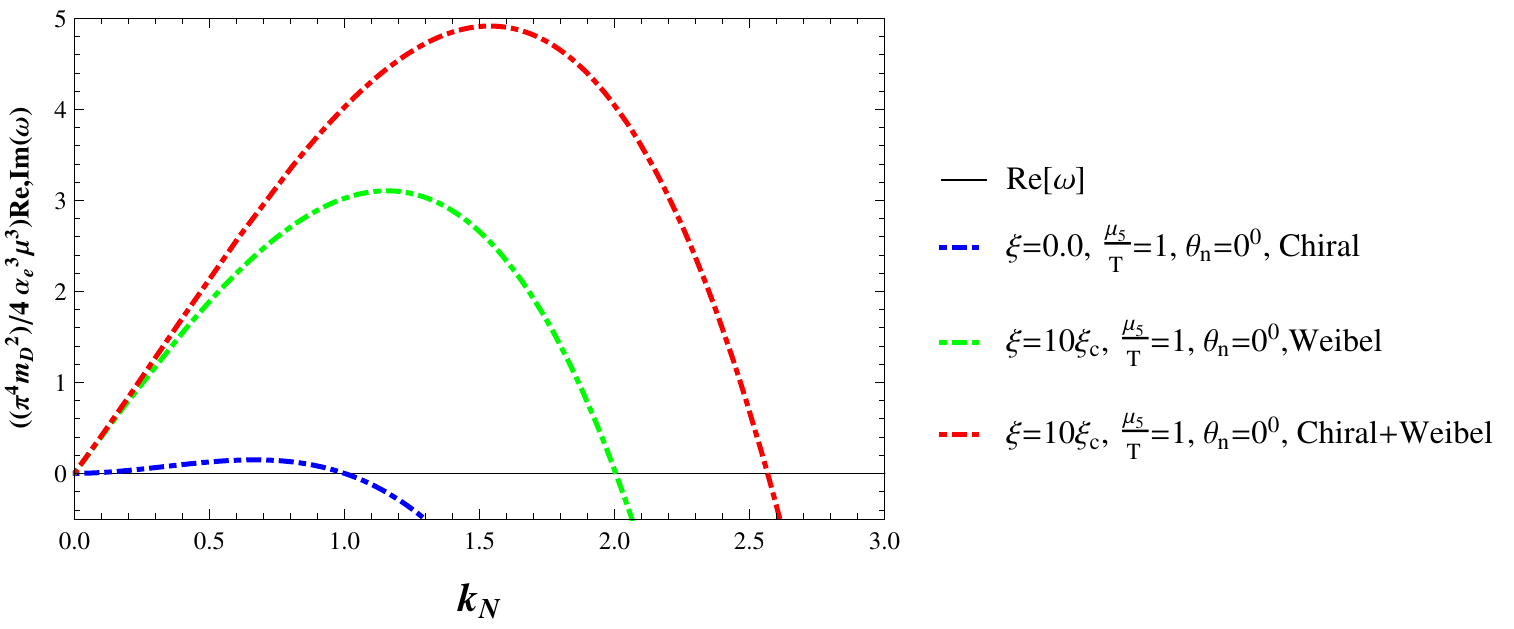}}
\subfigure[]{\includegraphics[bb=0 0 433 181,width=0.45\textwidth]{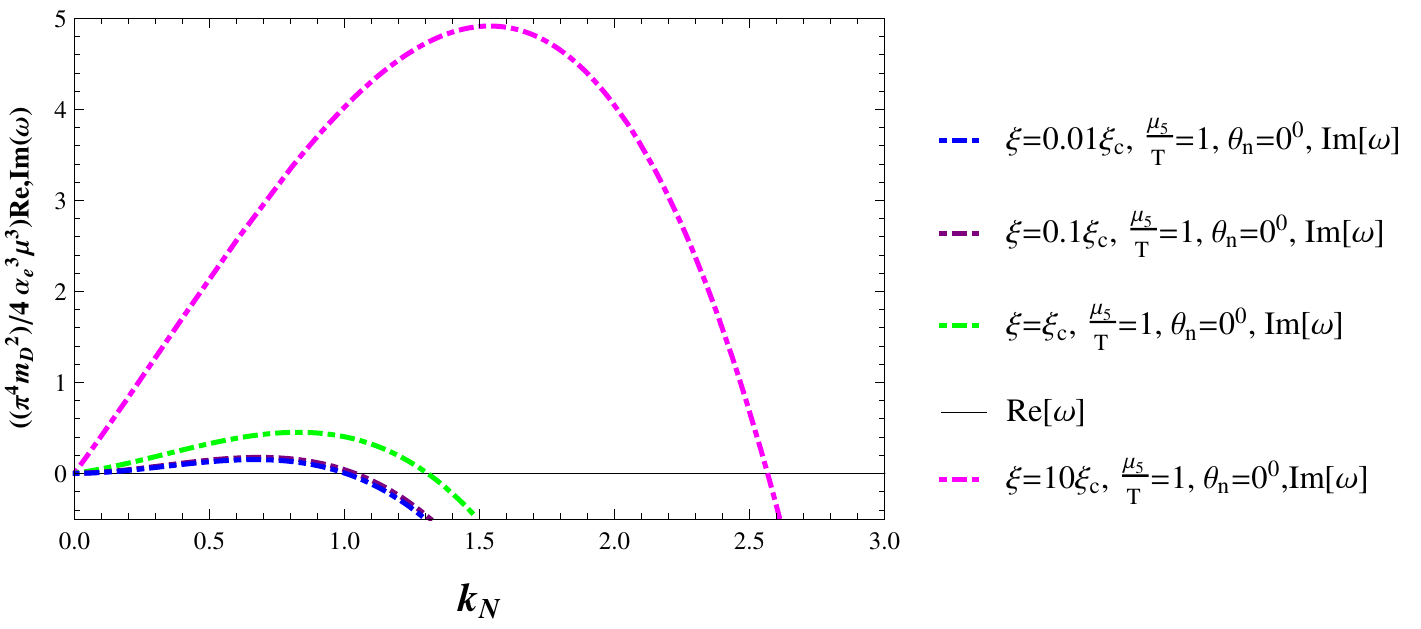}}
\subfigure[]{\includegraphics[bb=0 0 411 181,width=0.45\textwidth]{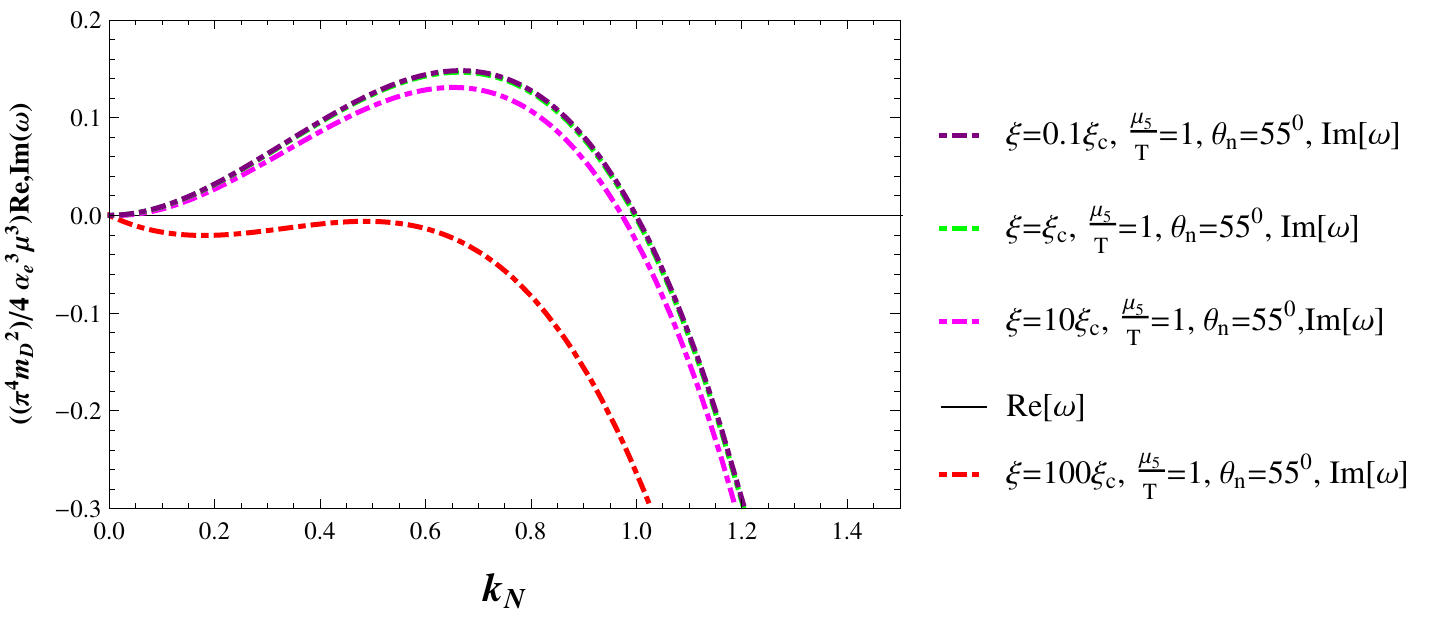}}

\end{center}
\end{figure}

\begin{figure}[H]
\begin{center}
\subfigure[]{\includegraphics[bb=0 0 411 181,width=0.45\textwidth]{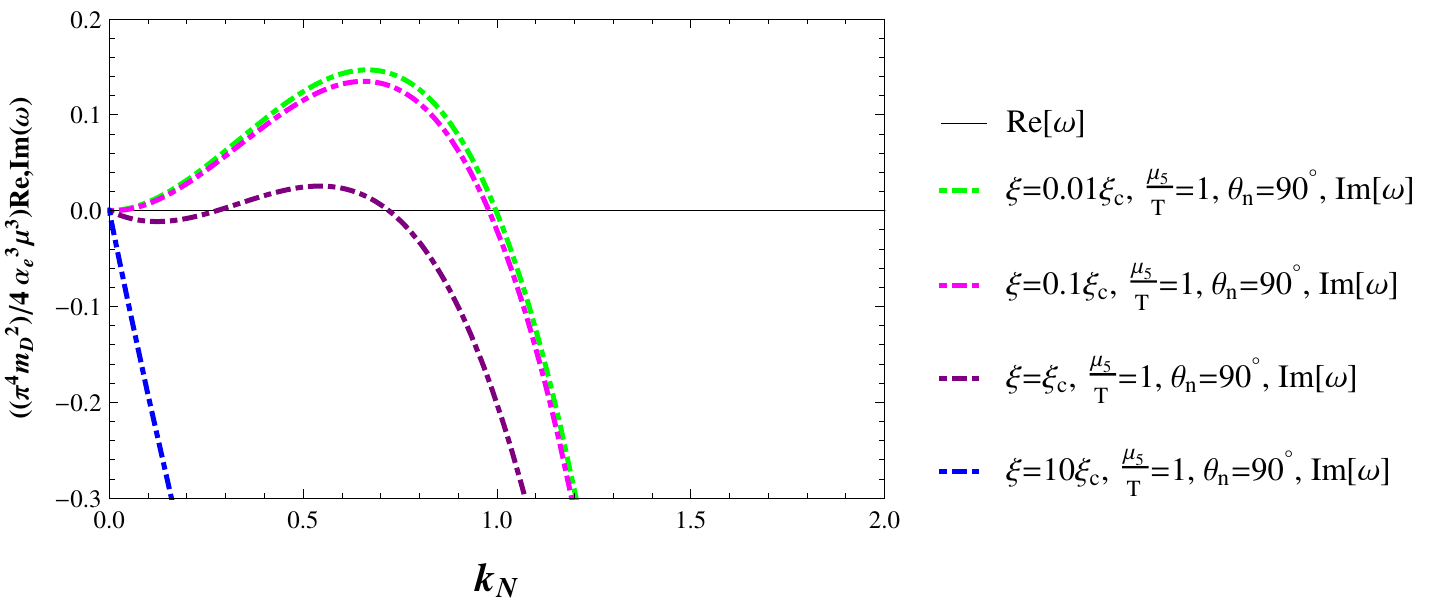}}
\subfigure[]{\includegraphics[bb=0 0 411 181,width=0.45\textwidth]{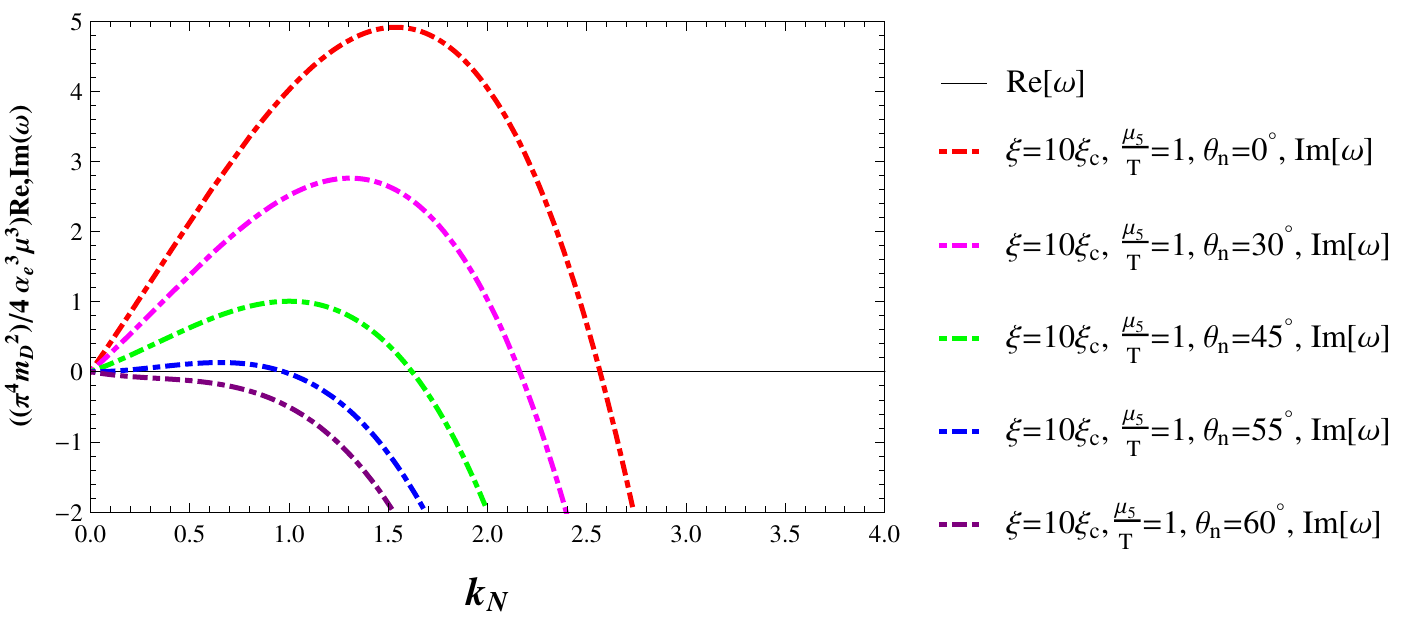}}
\subfigure[]{\includegraphics[bb=0 0 450 200,width=0.45\textwidth]{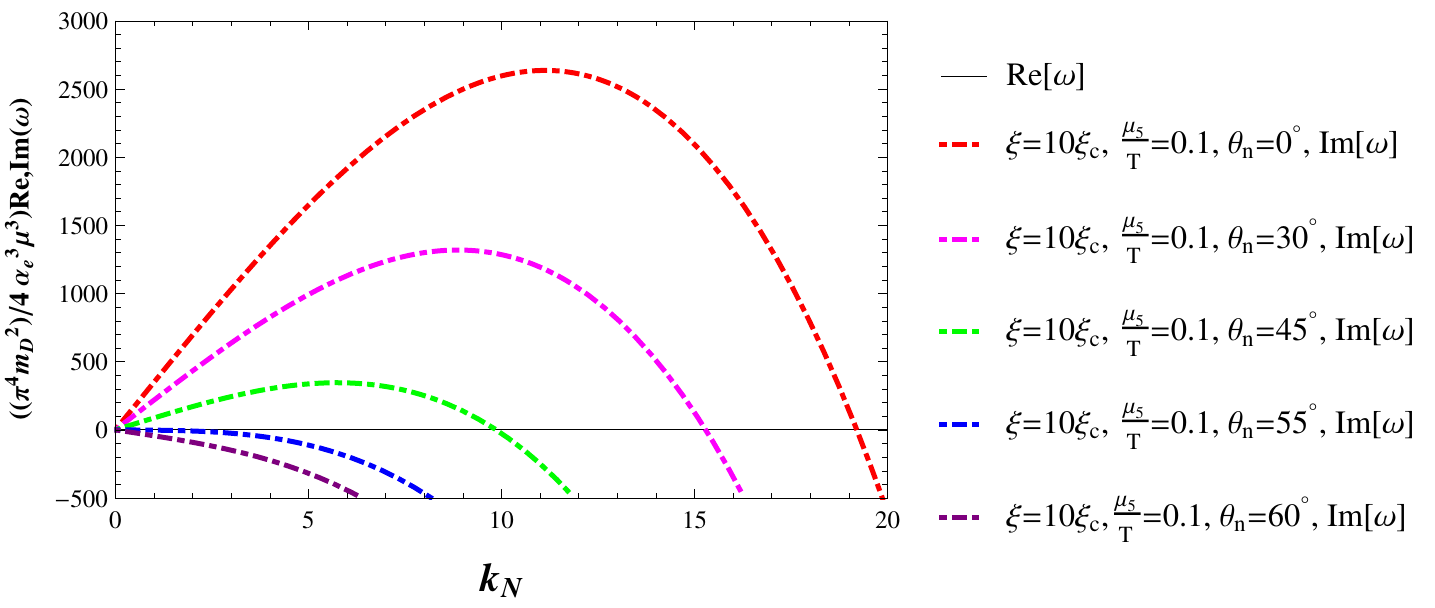}}
\subfigure[]{\includegraphics[bb=0 0 400 181,width=0.40\textwidth]{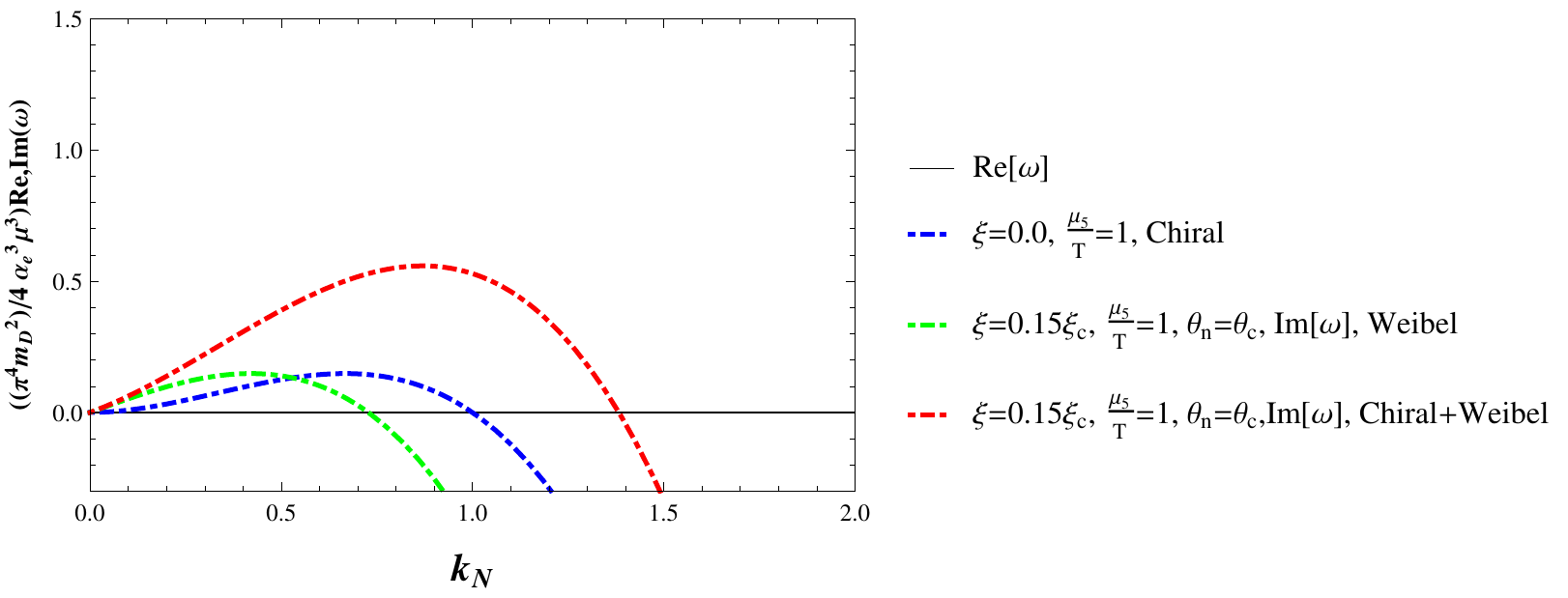}}
\caption 
 {\label{fig[1]} Shows plots of real and imaginary part of the dispersion relation. Here $\theta_{n}$ is the angle between 
  the wave vector $k$ and the anisotropy vector. Real part of dispersion relation is zero. Fig. (3a) 
  show plots for three cases: (i) Pure chiral (no anisotropy), (ii) Pure Weibel (chiral chemical potential=0) and (iii) 
  When both chiral and Weibel instabilities are present. Fig. (3b-3d) represent the case when both the 
  instabilities are present but the anisotropy parameter varies at different values of $\theta_{n}$ for fixed $\mu_5/T=1$.  Fig. (3e-3f) 
  represents the case when both instabilities are present for a fixed anisotropy parameter at different 
  values of $\theta_{n}$ when $\mu_5/T=1$ and $\mu_5/T=0.1$ respectively. Fig. (3g) represents the case when for a 
  particular value of $\theta_n\sim\theta_{c}$ both the instabilities have equal growth rates. Here frequency is normalized in unit
of $\omega/\left(\frac{4\alpha_{e}^3{\mu_5}^3}{\pi^{4}m^2_{D}}\right)$ and wave-number
$k$ by $k_{N}=\frac{\pi}{\mu_5\alpha_{e}}k$.} 
\label{fig[1]}
\end{center}
\end{figure}
In Fig.(3a) we have shown for $\theta_n=0$ the pure Weibel case ($\xi=10\xi_c$ and 
$\mu_5=0$) and the 
pure chiral-case ($\xi=0$ and $\mu_5 \neq 0$) with the case when both the instabilities
are present i.e. $\xi=10\xi_c$ and $\mu_5 \neq 0$. 
The plot shows that the pure Weibel modes
dominating over the pure chiral case. But the combined effect of both the instabilities is
much more pronounced. The maximum growth rate and the range of the instability are altered significantly
for the combined case.
In Figs. (3b-3d) we study the cases where both the instabilities are present
and $\xi$ and $\theta_n$ vary when $\mu_5/T=1$. {\it {It is important to note that in this analysis we are 
showing the plots of the dispersion relation following the same normazation as used in Ref.\cite{Akamatsu:2013} so 
that we can compare our results. Due to the normalization rescaling of dispersion relation for Weibel 
term picks up factor $\mu_5/T$ and therefore apart from $\xi$ and $\theta_n$ Weibel instability also becomes 
dependent on $\mu_5/T$. 
However, in order to take limit $\mu_5\rightarrow0$ one need to unscale normalized $Im[\omega]$ and $k_N$ in 
terms of $Im[\omega]$ and $k$.}}
Fig.(3b) shows clearly shows for $\theta_n=0$ 
when condition $\xi<<\xi_c$ is satisfied, the chiral instability
dominates over the Weibel modes. However, such values of $\xi$ are extremely small. For the cases when
$\xi\geq \xi_c$ the Weibel modes are dominating. Contribution from the Weibel modes is maximum for 
$\theta_n=0$ and the modes are strongly damped at $\theta_n=\pi/2$. Angular part in the dispersion
relation for the pure Weibel modes becomes zero when $\theta_n\approx 55^0$. In this case one can see
that chiral modes can remain dominant. This case is shown in Fig.(3c). It should be noted that for
the case when $\xi>>\xi_c$ the contribution from the coupling term between the Weibel and chiral
modes become sufficiently strong and it can again suppress the instability.
In Fig.(3d) we have shown the case when $\theta_n=\pi/2$. The modes with $\xi\geq \xi_c$
are strongly damped and there is no instability. Here the coupling term between the two modes
also contribute in the damping of the instability.
In Fig.(3e-3f) we have plotted the unstable modes for $\xi=10\xi_c$ for different values
of $\theta_n$ when $\mu_5/T=1$ and $0.1$ respectively. In this case see when $\mu_5/T=0.1$ i.e. ($\mu_5<<T$) 
the instability increases enormously. Now by comparing the growth rates of pure-Weibel and pure chiral modes,  
when $\mu_5/T=1$, one can find that they become 
equal at $ \theta_{c}
=\frac{1}{2}\cos^{-1}\left[\left(\frac{2}{27}\right)^{2/3}\frac{3\alpha_{e}}{\xi\pi^2}-\frac{1}{3}\right]
$. Fig.(3g) represents this case where we have shown that the growth rate of pure Weibel case at $\xi=0.15\xi_c$ 
becomes comparable to pure
chiral mode with $\xi=0$. The topmost (red) curve in this figure shows the case when both
the modes operate together. This case shows that the combined effect of the instability
can significantly alter the range and the growth rate of the instability.
 
In conclusion, we have studied collective modes in an anisotropic chiral plasma where the both Weibel
and chiral-imbalance instabilities are present. We have demonstrated that for $\theta_n=0$,
only for a very small values of the anisotropic parameter $\xi\sim\xi_c\ll1$ growth rates of the both instabilities
are comparable. For the cases when $\xi\geq 1$, $\xi<1$ but closer to unity and $\xi_c<\xi\ll1$,  
the Weibel modes dominate over the chiral-imbalance instability.
We have also shown for the case when $\xi\gg 1$, the chiral-imbalance can dominate over the Weibel modes for
certain values of $\theta_n$.

{\bf{Acknowledgements:}} 
The authors would like to thank the anonymous referees for their constructive comments and suggestions.


\begin{thebibliography}{unsrt}
	

\bibitem{Landau_kinetics}
 L.~D.~Landau and E.~M.~Lifshitz, {\it Physical Kinetics}
(Pergamon, New York, 1981).
%
\bibitem{Son:2009tf}
  D.~T.~Son and P.~Surowka,
  Phys.\ Rev.\ Lett.\  {\bf 103}, 191601 (2009)  [arXiv:0906.5044 [hep-th]].  

\bibitem{Banerjee:2012iz} 
  N.~Banerjee, J.~Bhattacharya, S.~Bhattacharyya, S.~Jain, S.~Minwalla, and T.~Sharma,
  JHEP {\bf 1209}, 046 (2012)
  [arXiv:1203.3544 [hep-th]].
  
\bibitem{Jensen:2012jy} 
  K.~Jensen,
  Phys.\ Rev.\ D {\bf 85}, 125017 (2012)
  [arXiv:1203.3599 [hep-th]].

\bibitem{Kharzeev:2010gr} 
  D.~E.~Kharzeev and D.~T.~Son,
  Phys.\ Rev.\ Lett.\  {\bf 106}, 062301 (2011).  [arXiv:1010.0038 [hep-ph]].  
\bibitem{Son:2012wh} 
  D.~T.~Son and N.~Yamamoto,
  Phys.\ Rev.\ Lett.\  {\bf 109}, 181602 (2012).
  [arXiv:1203.2697 [cond-mat.mes-hall]].
%
\bibitem{Zahed:2012}
I. Zahed, Phys. Rev. Lett. {\bf 109}, 091603 (2012).
%
\bibitem{Stephanov:2012}
M. A. Stephanov and Y. yin, Phys. Rev. Lett. {\bf 109}, 162001 (2012).\\
arXiv:1207.0747[hep-th]
%
\bibitem{Chen:2013}
Jiunn-Wei Chen, Shi Pu, Qun Wang and Xin-Nian Wang, Phys. Rev. Lett. {\bf 110}, 262301 (2013).
%
\bibitem{Loganayagam:2012}
R. Loganayagam and P. Surowka, J. High Energy Phs. {\bf 04}, 079 (2012).\\
arXiv:1201.2812[hep-th]
%
\bibitem{Xiao:2005}
D. Xiao, J. Shi and Q. Niu, Phys. Rev. Lett. {\bf 95}, 137204 (2005).
%
\bibitem{Berry}
  M.~V.~Berry, Proc.~R.~Soc.~Lond. A {\bf 392}, 45 (1984)
%
\bibitem{adler69}
S. Adler, Phys. Rev. {\bf 177}, 2426 (1969).
\bibitem{bell69}
J.S. Bell and R. Jackiw, Nuovo Cimento A {\bf 60} 4 (1969).
%
\bibitem{Nielsen:1983rb} 
  H.~B.~Nielsen and M.~Ninomiya,
  Phys.\ Lett.\ B {\bf 130}, 389 (1983).
%
\bibitem{vilenkin:80}
A. Vilenkin, Phys. Rev. D {\bf 22}, 3080 (1980).
%
\bibitem{Fukushima:2008xe} 
  K.~Fukushima, D.~E.~Kharzeev, and H.~J.~Warringa,
  Phys.\ Rev.\ D {\bf 78}, 074033 (2008).
%
\bibitem{Kharzeev08}
D. E. Kharzeev, L. D. McLerran and H. J. Warringa, Nucl. Phys. A {\bf 803}, 67 92007).
%
\bibitem{Warringa08}
H. J. Warringa, arXiv:0805.1384[hep-ph].
%
\bibitem{Abelev:2009}
B. I. Abelev {\it et. al.} [STAR Collaboration], Phys. Rev. Lett. {\bf 103}, 251601 (2009).\\
arXiv:0909.1739 [nucl-ex]
\bibitem{Abelev:2010}
B. I. Abelev {\it et. al.} [STAR Collaboration], Phys. Rev. C. {\bf 81}, 054908 (2010).\\
arXiv:0909.1717[nucl-ex].
%
\bibitem{Xiao:2010}
  D.~Xiao, M.-C.~Chang, and Q.~Niu, 
  Rev.\ Mod.\ Phys.\ {\bf 82}, 1959 (2010). 
  [arXiv:0907.2021 [cond-mat.mes-hall]].
%
\bibitem{kim13}
Heon-Jung Kim {\it et. al.}, Phys. Rev. Lett. {\bf 111}, 246603 (2013). 
%
\bibitem{Sasaki01}
K. Sasaki, arXiv:0106190 [cond-mat].
\bibitem{Son:2012zy} 
  D.~T.~Son and N.~Yamamoto,
  Phys.\  Rev.\ D {\bf 87}, 085016 (2013) [arxiv:1210.815].
\bibitem{Manuel:2013}
C. Manuel and J. M. Torres-Rincon, arXiv:1312.1158[hep-ph].
\bibitem{Itoyama:1983}
H. Itoyama and A. H. Mueller, Nucl. Phys. B, {\bf 165}, 349 (1983).
%
\bibitem{Liu:1988}
Y. Li and G. Ni, Phys. Rev. D. {\bf 38}, 3840 (1988).
%
\bibitem{Nicola:1994}
A. Gomez Nicola and R. F. Alvarez-Estrada, Int. J. Mod. Phys. A {\bf 9}, 1423 (1994).
\bibitem{Akamatsu:2013}
Y. Akamatsu and N. Yamamoto, Phys. Rev. Lett. {\bf 111}, 052002 (2013).
\bibitem{Nieves:1989}
J. F. Nieves and P. B. Pal, Phys. Rev. D {\bf 39}, 652 (1989).
\bibitem{Redlich:1985} 

  A.~N.~Redlich and L.~C.~R.~Wijewardhana,
  Phys.\ Rev.\ Lett.\  {\bf 54}, 970 (1985).

\bibitem{Rubakov:1985} 
  V.~A.~Rubakov,
  Prog.\ Theor.\ Phys.\  {\bf 75}, 366 (1986).

\bibitem{Joyce:1997} 
  M.~Joyce and M.~E.~Shaposhnikov,
  Phys.\ Rev.\ Lett.\  {\bf 79}, 1193 (1997).

\bibitem{Liane:2005} 
  M.~Laine,
  JHEP {\bf 0510}, 056 (2005).

\bibitem{Boyarsky:2012} 
  A.~Boyarsky, J.~Frohlich, and O.~Ruchayskiy,
  Phys.\ Rev.\ Lett.\  {\bf 108}, 031301 (2012).
%
\bibitem{Mro:1988}
S. Mr\'{o}wczynski, Phys. Lett. B {\bf 214}, 587 (1988),
S. Mr\'{o}wczynski, Phys. Lett. B {\bf 314}, 118 (1993), J. Randrup and S.Mr\'{o}wczynski,
arXiv:0303021[nucl-th].
%
\bibitem{Bhatt:1994}
J. R. Bhatt, P. K. Kaw and J. C. Parikh, Pramana- J. Phys. {\bf 43}, 467 (1994) 
%
\bibitem{Arnold:2003}
P. A. Arnold, J. Lenaghan and G. D. Moore,   J. High Energy Phs. {\bf 08}, 002 (2003).
%
\bibitem{Romatschke:2002}
P. Romatschke and R. Venugopalan, Phys. Rev. Lett. {\bf 96}, 062302 (2006)
%
\bibitem{romatschke}
Paul Romatschke and Michael Strickland, Phys. Rev. {\bf D 68}, 036004 (2003).
\bibitem{Tsitsadze:2009}
L. N. Tsintsadze, Phys. Plasmas, {\bf 16}, 094507 (2009).
\bibitem{Hu:1991}
B. Y. Hu and J. W. Wilkins, Phys. Rev. B {\bf 43}, 14 009 (1991).


%
\bibitem{Weibel:1959}
E.S. Weibel, Phys. Rev. Lett. {\bf2}, 83 (1959).
%
\bibitem{Fried:1959}
B.D. Fried, Phys. Fluids {\bf 2}, 337 (1959).
%
\bibitem{Krall}
N. A. Krall and A. W. Trivelpiece, {\it Principles of Plasma Physics}(San Francisco Press, San Francisco, 1986).

\bibitem{Kobes:1991}
R. Kobes, G. Kunstatter and A. Rebhan, Nucl. phys. B, {\bf 355} 1 (1991).


\end{thebibliography}
\end{document}